\documentclass[superscriptaddress,prl,preprint]{revtex4}
\begin{document}

\title{{\it In situ} epitaxial MgB$_2$ thin films for superconducting electronics}

\author{X. H. Zeng}

\affiliation{Department of Physics, The Pennsylvania State
University, University Park, PA 16802}

\author{A. V. Pogrebnyakov}
\affiliation{Department of Physics, The Pennsylvania State
University, University Park, PA 16802}

\author{A. Kotcharov}
\affiliation{Department of Physics, The Pennsylvania State
University, University Park, PA 16802}

\author{J. E. Jones}
\affiliation{Department of Physics, The Pennsylvania State
University, University Park, PA 16802}

\author{X. X. Xi}

\email [To whom correspondence should be addressed (at The
Pennsylvania State University). E-mail:] {xxx4@psu.edu}

\affiliation{Department of Physics, The Pennsylvania State
University, University Park, PA 16802}

\affiliation{Department of Materials Science and Engineering, The
Pennsylvania State University, University Park, PA 16802}

\affiliation{Materials Research Institute, The Pennsylvania State
University, University Park, PA 16802}

\author{E. M. Lysczek}
\affiliation{Department of Materials Science and Engineering, The
Pennsylvania State University, University Park, PA 16802}

\author{J. M. Redwing}
\affiliation{Department of Materials Science and Engineering, The
Pennsylvania State University, University Park, PA 16802}

\affiliation{Materials Research Institute, The Pennsylvania State
University, University Park, PA 16802}

\author{S. Y. Xu}

\affiliation{Department of Physics, The Pennsylvania State
University, University Park, PA 16802}

\author{Qi Li}

\affiliation{Department of Physics, The Pennsylvania State
University, University Park, PA 16802}

\affiliation{Materials Research Institute, The Pennsylvania State
University, University Park, PA 16802}

\author{J. Lettieri}
\affiliation{Department of Materials Science and Engineering, The
Pennsylvania State University, University Park, PA 16802}

\affiliation{Materials Research Institute, The Pennsylvania State
University, University Park, PA 16802}

\author{D. G. Schlom}
\affiliation{Department of Materials Science and Engineering, The
Pennsylvania State University, University Park, PA 16802}

\affiliation{Materials Research Institute, The Pennsylvania State
University, University Park, PA 16802}

\author{W. Tian}
\affiliation{Department of Materials Science and Engineering, The
University of Michigan, Ann Arbor, MI 48109}

\author{X. Q. Pan}
\affiliation{Department of Materials Science and Engineering, The
University of Michigan, Ann Arbor, MI 48109}

\author{Z. K. Liu}
\affiliation{Department of Materials Science and Engineering, The
Pennsylvania State University, University Park, PA 16802}

\affiliation{Materials Research Institute, The Pennsylvania State
University, University Park, PA 16802}

\begin{abstract}

A thin film technology compatible with multilayer device
fabrication is critical for exploring the potential of the 39-K
superconductor magnesium diboride for superconducting
electronics. Using a Hybrid Physical-Chemical Vapor Deposition
(HPCVD) process, it is shown that the high Mg vapor pressure
necessary to keep the MgB$_2$ phase thermodynamically stable can
be achieved for the {\it in situ} growth of MgB$_2$ thin films.
The films grow epitaxially on (0001) sapphire and (0001) 4H-SiC
substrates and show a bulk-like $T_c$ of 39\,K, a $J_c$(4.2K) of
$1.2 \times 10^7$ A/cm$^2$ in zero field, and a $H_{c2}(0)$ of
29.2 T in parallel magnetic field. The surface is smooth with a
root-mean-square roughness of 2.5 nm for MgB$_2$ films on SiC.
This deposition method opens tremendous opportunities for
superconducting electronics using MgB$_2$.

\end{abstract}

\maketitle

The newly-discovered 39-K superconductor magnesium diboride
\cite{Nagamatsu01} holds great promises for superconducting
electronics. However, success in fabricating MgB$_2$ Josephson
junctions, the fundamental element of superconducting
electronics, has been very limited due to the lack of an adequate
thin film technology \cite{Mijatovic02}. The main difficulty in
depositing MgB$_2$ thin films is that a very high Mg vapor
pressure is necessary for the thermodynamic stability of the
MgB$_2$ phase at elevated temperatures \cite{ZKLiu01}. The films
deposited by the current technologies either have reduced $T_c$
and poor crystallinity
\cite{Blank01,Christen01,Shinde01,Grassano01,Zeng01,Ueda01,Jo02},
or require {\em ex situ} annealing in Mg vapor
\cite{WNKang01,HYZhai01,Eom01}. The surfaces of these films are
rough and non-stoichiometric \cite{HYZhai01,Mijatovic02,SYLee02},
undesirable for Josephson junction devices. In this report, we
show that epitaxial MgB$_2$ thin films can be deposited {\it in
situ} by Hybrid Physical-Chemical Vapor Deposition (HPCVD), a
novel combination of Physical Vapor Deposition (PVD) and Chemical
Vapor Deposition (CVD). At a substrate temperature of around
750$^{\circ}$C, the films grow epitaxially on (0001) sapphire and
(0001) 4H-SiC substrates. They show bulk-like $T_c$, high critical
current density $J_c$, and high upper critical field $H_{c2}$. The
surfaces are smooth. The process is compatible with multilayer
device fabrication and easy to scale up. These results mark the
removal of a major road block towards a reliable MgB$_2$ Josephson
junction technology.

MgB$_2$ has many properties that make it very attractive for
superconducting electronics. Unlike cuprate high-temperature
superconductors, MgB$_2$ seems to be a phonon-mediated
superconductor \cite{Budko01} with a relatively long coherence
length \cite{Finnemore01}. The grain boundaries in MgB$_2$ are
not weak links \cite{Larbalestier01,Finnemore01}. From the device
perspective, MgB$_2$ is similar to Nb but with a higher $T_c$ and
larger energy gaps, even though two energy gaps exist
\cite{Tsuda01,Schmidt02}. This will lead to higher speeds and
higher operating temperatures than the Nb-based superconducting
digital circuits. Moving the operating temperature from 5 K for
Nb to 20 K for MgB$_2$ greatly reduces the cryogenic
requirements, thus making superconducting electronics much more
competitive.

For the deposition of MgB$_2$ thin films, a thermodynamic analysis
predicted a growth window in the pressure-temperature phase
diagram in which the MgB$_2$ phase is thermodynamically stable
\cite{ZKLiu01}. At temperatures necessary for epitaxial growth,
the Mg vapor pressure in this growth window is very high for many
vacuum deposition techniques. In practice, {\it in situ} growth
of MgB$_2$ thin films has been demonstrated below 320$^{\circ}$C
where the necessary Mg vapor pressure for phase stability is low
\cite{Ueda01,Jo02}. However, the temperature is too low for
epitaxial growth, and the resultant films have lower $T_c$
(34\,K) and poor crystallinity. An alternative technique employs
annealing of Mg-B or Mg-MgB$_2$ mixtures {\it in situ} in the
growth chamber, at temperatures and duration such that severe Mg
loss or MgB$_2$ decomposition does not occur
\cite{Blank01,Christen01,Shinde01,Grassano01,Zeng01,Newman02}.
Films produced by such techniques also show lower $T_c$ (34\,K)
and poor crystallinity \cite{Zeng01}. By annealing B films {\it
ex situ} in high Mg vapor pressure, temperatures as high as
900$^{\circ}$C can be used \cite{WNKang01,HYZhai01,Eom01}. The
MgB$_2$ films thus produced show bulk $T_c$ (39\,K) and epitaxy
\cite{SDBu02,WTian02}. Unfortunately, this two-step {\it ex situ}
annealing process is not desirable for multilayer device
fabrication.

Our approach to meet the challenge of generating a high Mg vapor
pressure is HPCVD. It is in principle similar to the Physical
Vapor Transport (PVT) technique used to grow semiconductor single
crystals \cite{Kamler00}. In HPCVD, a standard vertical CVD
quartz reactor is used and diborane, B$_2$H$_6$, is used as the
boron precursor gas. Unlike conventional CVD which utilizes
gaseous sources only, HPCVD uses heated bulk Mg (99.95\%) as the
Mg source. Before the deposition, a single crystal substrate is
placed on the top surface of a SiC-coated graphite susceptor and
bulk pieces of Mg are placed nearby. The reactor is then purged
with purified N$_2$ gas and purified H$_2$ gas. The susceptor,
along with the substrate and Mg pieces, are heated by an
inductance heater to 730-760$^{\circ}$C in the H$_2$ ambient. A
mixture of 1000 ppm B$_2$H$_6$ in H$_2$ is then introduced into
the reactor to initiate growth.  The total pressure in the
reactor is maintained at 100-700 Torr throughout the process. Due
to the relatively high gas pressure and the flow pattern of the
carrier gas in the reactor, the vapor from the heated pure Mg
results in a high Mg vapor pressure near the substrate. When the
B$_2$H$_6$ gas is not flowing through the reactor, there is no
film deposition because of the low sticking coefficient of Mg at
high temperatures \cite{Newman02}. Once the B$_2$H$_6$ gas begins
to flow, a MgB$_2$ film starts to grow on the substrate. The
deposition rate depends on the flow rate of B$_2$H$_6$ gas. For
example, a flow rate of 25 sccm for the 1000 ppm B$_2$H$_6$
mixture results in a deposition rate of about 2-3 \AA/s. After
growth, the B$_2$H$_6$ gas is switched out of the reactor and the
sample is cooled in the H$_2$ carrier gas to room temperature.

Films were deposited on (0001) sapphire and (0001) 4H-SiC
substrates. They were characterized for phase and structural
information using four-circle x-ray diffraction with
CuK$_{\alpha}$ radiation and a graphite monochromator. Fig. 1(A)
shows a $\theta-2\theta$ scan of a MgB$_2$ film on a (0001)
sapphire substrate. Beside the substrate peaks the only peaks
observed arise from the (0001) planes of MgB$_2$, indicating that
a phase-pure MgB$_2$ film with $c$-axis orientation is obtained.
The $c$ lattice constant is 3.512$\pm$0.005 \AA, in agreement with
the bulk value \cite{Budko01}. The full widths at half maximum
(FWHM) of the 0002 peak were 0.47$^{\circ}$ and 1.6$^{\circ}$ in
2$\theta$ and $\omega$, respectively. The in-plane orientation
was probed with a $\phi$-scan of the 10$\overline{1}$1 MgB$_2$
reflection shown in Fig. 1(B). A six-fold symmetry characteristic
of a (0001) oriented MgB$_2$ film with in-plane epitaxy is seen.
The $a$ lattice constant is 3.08$\pm$0.02 \AA, smaller than the
bulk value of 3.14 \AA\,\cite{Budko01}. The FWHM of this peak in
$\phi$ is $1.65^{\circ}$. The plot reveals small peaks at
30$^{\circ} \pm n$ 60$^{\circ}$ (where $n$ is an integer),
indicating minimal amounts of 30$^{\circ}$ rotational twinning.
The dominant in-plane epitaxial relationship is that the
hexagonal MgB$_2$ lattice is rotated by 30$^{\circ}$ (i.e.
[10$\overline{1}$0] MgB$_2 \parallel [2\overline{1}\overline{1}$0]
Al$_2$O$_3$) to match the hexagonal lattice of sapphire, which has
a lattice constant of 4.765 \AA. X-ray diffraction of MgB$_2$
films on (0001) SiC substrates also revealed epitaxial growth
with (0001) MgB$_2 \parallel$ (0001) SiC and
[2$\overline{1}\overline{1}$0] MgB$_2 \parallel
[2\overline{1}\overline{1}$0] SiC and without 30$^{\circ}$
rotational twinning. The lattice parameters obtained are $c=3.516
\pm 0.005$ \AA\,and $a=3.09 \pm 0.02$ \AA, respectively. Due to
the close lattice match between SiC ($a=3.07$ \AA\,for 4H-SiC)
and MgB$_2$, the hexagonal lattice of MgB$_2$ grows directly on
top of the hexagonal lattice of SiC.

The same film described in Fig. 1 was studied by cross-section
transmission electron microscopy (TEM). The measurement was
performed in a JEOL 4000 EX microscope operated at 400 kV,
providing a point-to-point resolution of 0.17 nm. Fig. 2(A) shows
a low magnification bright-field TEM image. Fig. 2(B) and (C) are
selected-area electron diffraction patterns taken from the film
and the substrate with the incident electron beam along the same
direction. Fig. 2(D) shows a high-resolution TEM image of the
film/substrate interface. The results confirm the epitaxial
growth of a $c$-axis oriented MgB$_2$ film on the sapphire
substrate. The diffraction pattern in Fig. 2(B) is from the
[11$\overline{2}$0] zone axis of MgB$_2$; Fig. 2(C) is from the
[10$\overline{1}$0] zone axis of sapphire. This corroborates the
x-ray diffraction results on the in-plane epitaxial relationship
between the film and the sapphire substrate. Cross-section and
plane-view TEM observations also show that the film consists of
strongly faceted columns with different heights, resulting from
island growth. These columns have hexagonal shapes with very
smooth top surfaces, and they have the same crystallographic
orientations. An intermediate layer at the MgB$_2$/Al$_2$O$_3$
interface is seen in the high-resolution image. Fourier-transform
studies and nano-sized probe electron diffraction reveal that the
image characteristics of this layer result from a superposition
of MgO regions with the epitaxially-grown MgB$_2$. This
intermediate layer is much thinner and simpler than that in the
{\it ex situ} annealed epitaxial MgB$_2$ films, where layers of
MgAl$_2$O$_4$ and MgO are observed \cite{WTian02}.

It should be noted that the x-ray diffraction spectra are free of
MgO peaks, which commonly appear in MgB$_2$ thin films grown by
other techniques \cite{WNKang01,Eom01,Christen01,Blank01,Zeng01}.
In cross-section TEM, MgO regions are seen only in a very thin
layer near the MgB$_2$/Al$_2$O$_3$ interface, which may be the
result of oxygen diffusion out of the substrate. We attribute the
minimal oxygen contamination in the bulk of the MgB$_2$ films to
the reducing ambient resulting from the use of hydrogen as the
carrier gas in the process. When the susceptor and bulk Mg are
heated to 750$^{\circ}$C, H$_2$ effectively eliminates the oxide
on the Mg surface and suppresses MgO formation during growth.

The superconducting and transport properties of the films were
characterized by resistivity measurement using the standard
four-point method. Fig. 3(A) shows a resistivity vs. temperature
curve for a 2000\,\AA\,MgB$_2$ film on a sapphire substrate. The
inset shows the details near the superconducting transition, and
a zero-resistance $T_c$ of 39.3\,K is observed. The $T_c$ value of
39\,K is the same as that in the bulk materials, and has been
repeatedly obtained in our MgB$_2$ films on sapphire and SiC
substrates. The resistivity of the film is 10.5 $\mu \Omega$cm at
300\,K and 2.8 $\mu \Omega$cm before the superconducting
transition, giving a residual resistance ratio
$RRR=R$(300K)/$R$(40K) of 3.7. In our films with $T_c \sim$
39\,K, we normally find $RRR$ values of around 3.

The transport $J_c$ for a 2900 \AA\,thick MgB$_2$ films on a
sapphire substrate is shown in Fig. 3(B) as a function of
temperature and magnetic field. It was measured on a 30-$\mu$m
wide bridge using a Quantum Design PPMS system with a 9-T
superconducting magnet. The zero-field $J_c$ is $1.2 \times 10^7$
A/cm$^2$ at 4.2 K, comparable to the best reported value in {\it
ex situ} annealed films \cite{Eom01,HJKim01}. The $J_c$ drops
under applied magnetic fields. When $H\perp$ film the rates at
which $J_c$ decreases is similar to that in {\it ex situ}
annealed films with less oxygen contamination
\cite{Eom01,HJKim01}, but faster than in {\it ex situ} annealed
films with substantial oxygen contamination \cite{Eom01}. This is
consistent with the suggestion by Eom {\it et al.} that oxygen
contamination in MgB$_2$ films provides pinning centers
\cite{Eom01}. Because there is minimal oxygen contamination in
our films as shown by the x-ray diffraction and TEM, the vortex
pinning is weaker than in oxygen contaminated films. The
suppression of $J_c$ when $H\parallel$ film is much slower than
for $H\perp$ film.

High-quality epitaxial MgB$_2$ thin films are valuable for basic
studies of physical properties such as electronic anisotropy
\cite{Jung01,Ferdeghini01,Patnaik01}. Fig. 3(C) shows the
temperature dependence of the upper critical fields of a MgB$_2$
film on a SiC substrate when the magnetic field is applied
perpendicular ($H_{c2}^{\perp}(T)$, solid triangles) and parallel
($H_{c2}^{\parallel}(T)$, open triangles) to the film plane (the
(0001) plane). They are determined from the resistive transition
by the intersection of a line tangent to the transition midpoint
with a linear extrapolation of the normal state resistivity. A
clear anisotropy between $H_{c2}^{\parallel}(T)$ and
$H_{c2}^{\perp}(T)$ is observed. Because of the upward curvature
near $T_c$, we calculate d$H_{c2}/$d$T$ using the
lower-temperature part of the data. Using the expression for the
dirty-limit type-II superconductors, $H_{c2}(0) = 0.69
T_c$d$H_{c2}/$d$T$, we find $H_{c2}^{\parallel}(0)$ and
$H_{c2}^{\perp}(0)$ to be 29.2 T and 23.2 T, respectively, which
yields an anisotropy ratio, $\eta \equiv
H_{c2}^{\parallel}(0)/H_{c2}^{\perp}(0)$ of 1.26. Both the
$H_{c2}$ and $\eta$ values are similar to those in {\it ex situ}
annealed MgB$_2$ films \cite{Jung01}.

The excellent epitaxy and superconducting properties of the
MgB$_2$ films demonstrate that {\it in situ} growth of
high-quality epitaxial MgB$_2$ films is possible as long as a
sufficiently high Mg vapor pressure is produced. At
750$^{\circ}$C, this pressure is about 10 mTorr in the
thermodynamic growth window \cite{ZKLiu01}. The HPCVD technique
successfully achieves this value. The results also demonstrate
the automatic composition control in the adsorption-controlled
growth of films containing volatile species \cite{JYTsao93}.
During the growth, the Mg:B ratio arriving at the substrate is
likely above the 1:2 stoichiometry, but the extra Mg remains in
the gas phase and the desired MgB$_2$ phase is always obtained.
The deposition rate of MgB$_2$ is determined by the influx of
diborane.

The {\it in situ} deposited MgB$_2$ films have mirror-like shiny
surfaces. We have characterized the film surface by atomic force
microscopy (AFM) (Nanoscope III). In Fig. 4, AFM images of
MgB$_2$ films on (A) sapphire and (B) SiC substrates are shown.
Clearly, films on the two substrates have very different
morphologies. The surface of the film on SiC is smoother than that
on sapphire substrate. When measured over a large scan area, for
example $10 \times 10 \mu$m$^2$, the root-mean-square (RMS)
roughness for sapphire and SiC substrates are similar, $\sim$ 4
nm. However, on a $1 \times 1 \mu$m$^2$ scale, the RMS roughness
is $\sim$ 4 nm for the film on sapphire and $\sim$ 2.5 nm for the
film on SiC. This is much smoother than any MgB$_2$ film surface
published so far \cite{HYZhai01,Ferdeghini01}. The AFM images
also show that the majority of the growth columns have dimensions
larger than 100 nanometers, which is consistent with the TEM
result.

The results presented here demonstrate that HPCVD is a viable film
deposition technique for MgB$_2$ superconducting electronics. The
epitaxial films produced by HPCVD have excellent superconducting
properties and the small roughness of the film is promising for
Josephson junctions. Furthermore, the {\it in situ} process can
be readily scaled up to deposit over large substrate areas and
offers the potential for multilayer heterostructure fabrication.
As the technique is further refined and the film quality
improved, emphasis can shift from MgB$_2$ film deposition to the
fabrication of MgB$_2$ Josephson junctions and circuits. Finally,
the availability of high-quality epitaxial thin films is
anticipated to benefit a variety of fundamental studies of
MgB$_2$.

\newpage
{\bf{References and Notes}}

\noindent [29] We gratefully acknowledge Arsen Soukiassian,
Srinivasan Raghavan, and Kok-Keong Lew for their help in setting
up the HPCVD system. This work is supported in part by ONR under
grant Nos. N00014-00-1-0294 (XXX) and N0014-01-1-0006 (JMR), by
NSF under grant Nos. DMR-9876266 and DMR-9972973 (QL),
DMR-9875405 and DMR-9871177 (XQP), DMR-9983532 (ZKL), and by DOE
through grant DE-FG02-97ER45638 (DGS).

\begin{figure}

FIGURE CAPTIONS

\caption{X-ray diffraction spectra of a MgB$_2$ film on a (0001)
sapphire substrate showing an epitaxial orientation relationship.
(A) The $\theta-2\theta$ scan showing a phase-pure $c$-axis
oriented MgB$_2$ film. The spectrum is free from MgO peaks. (B)
The $\phi$-scan of the 10$\overline{1}$1 MgB$_2$ reflection. The
six-fold symmetry indicates a (0001)-oriented MgB$_2$ film with
in-plane epitaxy. Weak peaks at 30$^{\circ} \pm n$ 60$^{\circ}$
indicate minimal amounts of 30$^{\circ}$ rotational twinning. The
dominant epitaxial relationship is that the hexagonal MgB$_2$
lattice is rotated by 30$^{\circ}$ to match the hexagonal lattice
of sapphire. For (0001) 4H-SiC substrate, MgB$_2$ films grow
epitaxially with $c$-axis orientation and the hexagonal lattice
of MgB$_2$ grows directly on top of the hexagonal lattice of SiC.}

\caption{Microstructure and interfacial atomic structure of the
same film described in Fig. 1 studied by cross-sectional TEM. (A)
A low magnification bright-field TEM image. (B) Selected-area
electron diffraction (SAED) pattern from the MgB$_2$ film showing
a [11$\overline{2}$0] zone-axis diffraction pattern. (C) An SAED
pattern from the sapphire substrate with the incident electron
beam along the same direction as in (B), showing a
[10$\overline{1}$0] zone-axis diffraction pattern. The results
confirm the in-plane epitaxial relationship described in Fig. 1.
(D) A high-resolution TEM image of the film/substrate interface
showing an intermediate layer. The image characteristics of the
intermediate layer results from a superposition of epitaxial
MgB$_2$ and regions of MgO.}

\caption{Superconducting and transport properties of MgB$_2$ thin
films. (A) Resistivity vs. temperature for a 2000\,\AA\,thick
MgB$_2$ film on a sapphire substrate. A zero-resistance $T_c$ of
39.3\,K is clearly shown in the inset. (B) Critical current
density $J_c$ of a 2900\,\AA\,thick MgB$_2$ film on a sapphire
substrate as a function of magnetic field at different
temperatures. The 4.2 K data include both $H \parallel$ film and
$H \perp$ film, and only $H \perp$ film data are shown for other
temperature. (C) Upper critical fields $H_{c2}$ of a MgB$_2$ film
on a SiC substrate when the magnetic field is applied
perpendicular (solid triangles) and parallel (open triangles) to
the film plane. We find $H_{c2}^{\parallel}(0) \approx 29.2$ T,
$H_{c2}^{\perp}(0) \approx 23.2$ T, and the anisotropy ratio
$\eta = 1.26$.}
\end{figure}

\begin{figure}
\caption{AFM images of epitaxial MgB$_2$ films on (A) a sapphire
and (B) a SiC substrates. The vertical scale is 200 nm/div. The
two films have very different surface morphologies, and the film
on SiC is smoother than that on the sapphire substrate. The RMS
roughness is $\sim$ 4 nm for the film on sapphire and $\sim$ 2.5
nm for the film on SiC. The AFM images also show that the
majority of the growth columns have dimensions larger than 100
nanometers.}

\end{figure}

\end{document}